\documentclass[showpacs, oneside, twocolumn, prd, amsmath, amssymb, nofootinbib, superscriptaddress]{revtex4-1}

\usepackage{cases}
\usepackage{amsmath}
\usepackage{amssymb}
\usepackage{amsfonts}
\usepackage{amssymb}
\usepackage{dcolumn}
\usepackage{bm}
\usepackage{bbm}
\usepackage{graphicx}
\usepackage{xcolor}
\usepackage{array}
\usepackage{subfigure}
\usepackage{hyperref}
\usepackage{wasysym}
\usepackage{mathrsfs}
\usepackage{bm}
\usepackage{array}

\definecolor{gr}{rgb}{0.6,0.6,0.6}

\newcommand{\be}{\begin{equation}}
\newcommand{\ee}{\end{equation}}
\newcommand{\ba}{\begin{eqnarray}}
\newcommand{\ea}{\end{eqnarray}}

\newcommand{\gsim}{\mathrel{\hbox{\rlap{\lower.55ex \hbox {$\sim$}}
			\kern-.3em \raise.4ex \hbox{$>$}}}}
\newcommand{\lsim}{\mathrel{\hbox{\rlap{\lower.55ex \hbox {$\sim$}}
			\kern-.3em \raise.4ex \hbox{$<$}}}}

\hypersetup{colorlinks=true,
breaklinks=true,
pdfstartview=Fit,
linkcolor=blue,
citecolor=blue,
urlcolor=blue}

\begin{document}

\title{Updated constraints on superconducting cosmic strings from the astronomy of fast radio bursts}

\author{Batool Imtiaz}
%\email{batool24@mail.ustc.edu.cn}
\affiliation{Department of Astronomy, School of Physical Sciences, University of Science and Technology of China, Hefei, Anhui 230026, China}
\affiliation{CAS Key Laboratory for Researches in Galaxies and Cosmology, University of Science and Technology of China, Hefei, Anhui 230026, China}
\affiliation{School of Astronomy and Space Science, University of Science and Technology of China, Hefei, Anhui 230026, China}

\author{Rui Shi}
%\email{rshi9@jh.edu}
\affiliation{School of Astronomy and Space Science, University of Science and Technology of China, Hefei, Anhui 230026, China}
\affiliation{Department of Physics and Astronomy, Johns Hopkins University, Baltimore, Maryland 21218, USA}

\author{Yi-Fu Cai}
\email{yifucai@ustc.edu.cn}
\affiliation{Department of Astronomy, School of Physical Sciences, University of Science and Technology of China, Hefei, Anhui 230026, China}
\affiliation{CAS Key Laboratory for Researches in Galaxies and Cosmology, University of Science and Technology of China, Hefei, Anhui 230026, China}
\affiliation{School of Astronomy and Space Science, University of Science and Technology of China, Hefei, Anhui 230026, China}

\begin{abstract}	
In this article we update constraints on superconducting cosmic strings (SCSs) in light of the recent observational developments of fast radio bursts (FRBs) astronomy. Assuming strings follow an exponential distribution characterized by current, we show that two parameters in our context, which are the characteristic tension ($G\mu$) and a parameter which describes the aforementioned exponential distribution ($I_c$), can be constrained by FRB experiments. Particularly, we investigate data sets from Parkes and ASKAP. We looked through a parameter space where $G\mu \sim [10^{-17}, 10^{-12}]$ and $I_c\sim [10^{-1}, 10^{2}]$ GeV, and found our results show that Parkes jointly with ASKAP can constrain the parameter space for SCSs.
\end{abstract}

%\pacs{98.80.-k, 95.36.+x, 04.50.Kd, 04.30.Nk}

\maketitle

\section{Introduction}

Since the first observation of the fast radio transient phenomenon in the universe \cite{Lorimer:2007qn}, the astronomy of fast radio bursts (FRBs) has attracted broad interests from both the observational and theoretical perspectives \cite{Petroff:2019tty, Cordes:2019cmq}. Due to their large dispersion measure and bright pulses, many researchers attempt to investigate these mysterious signals from cosmological interpretations. A major effort has been made to establish the broad-band, wide-field surveys for the purpose of hunting for FRB events from the perspective of observational astronomy, such as Parkes \cite{Keane:2011mj, Thornton:2013iua, Champion:2015pmj, Bhandari:2017qrj}, Arecibo \cite{Spitler:2014fla, Patel:2018ubs}, Green Bank Telescope (GBT) \cite{Masui:2015kmb}, Square Kilometre Array (SKA) \cite{Bannister:2017sie, Macquart:2018rsa}, Molonglo Observatory Synthesis Telescope (MOST) \cite{Caleb:2017vbk, Farah:2018buz}, and Canadian Hydrogen Intensity Mapping Experiment (CHIME) \cite{Amiri:2019qbv}. Meanwhile, different theoretical models have been proposed to understand the possible origin of these FRB signals \cite{Platts:2018hiy}, namely, synchrotron maser emission from young magnetars in supernova remnants \cite{Metzger:2017wdz}, radiation from cosmic string cusps \cite{Brandenberger:2017uwo}, charged primordial black hole binaries coalescence \cite{Deng:2018wmy}, and many other possible models \cite{Wang:2016dgs, Dai:2016qem, Zhang:2017zse, Palaniswamy:2017aze}. Finding the origin of FRBs has become one of the most important tasks for understanding them.

Cosmic strings \cite{Vilenkin:2000jqa}, being one-dimensional topological defects, formed during the early evolution of the Universe \cite{Kibble:1976sj}. It might be possible that cosmic strings are superconducting wires if they carry electric charge \cite{Witten:1984eb}. These so-called superconducting cosmic strings (SCSs) can be achieved by introducing a charged scalar field whose flux is trapped inside strings with the electromagnetic gauge invariance broken, and hence can yield electromagnetic effects \cite{Vilenkin:2000jqa}. These primordial relics, if exist in the sky, could behave as giant wires that may release electromagnetic signals in a wide range of frequencies \cite{Vilenkin:1986zz, Garfinkle:1987yw}. Thus, it is natural to explore the hypothetical possibility for SCSs to explain the observed FRB events, namely, due to the oscillations of string loops \cite{Vachaspati:2008su}, dynamics of string cusps and kinks \cite{Cai:2011bi, Cai:2012zd, Ye:2017lqn}, and effects of the magnetic field upon string loops \cite{Yu:2014gea}. 

The parameters of SCSs (string tension $G\mu$ and current on string $I$) have been constrained by different types of astronomical observations. Namely, the CMB analyses based on the Wilkinson Microwave Anisotropy Probe (WMAP) and the South Pole Telescope can lead to an upper bound on the string tension of $G\mu < 1.7 \times 10^{-7}$ \cite{Dvorkin:2011aj}, and this bound was later improved to $G\mu < 1.3 \times 10^{-7}$ with the data from the Planck satellite \cite{Ade:2013xla}. Both CMB and the pulsar timing measurements put a robust bound on the string tension in the same order \cite{vanHaasteren:2011ni, Pshirkov:2009vb, Tuntsov:2010fu, Olmez:2010bi, Binetruy:2012ze, Sanidas:2012ee}. The spectral distortions of the CMB photons \cite{Sanchez:1988ek, Tashiro:2012nb, Acharya:2019xla} suggested some additional constraints on parameter space i.e. $10^{-19} < G\mu < 10^{-7}$ and $I > 10^4 ~ {\rm GeV}$ would be excluded \cite{Kogut:2011xw}. In most of the previous works, it has been assumed that all SCSs take the exact same value of the current, which would greatly simplify the analysis while one could grasp the basic picture of the underlying physics. However, in a much realistic situation, one would expect a probability distribution for the currents inside various strings in the universe. To explore this possibility, we in this article put forward a novel parameterization for SCSs, in which the currents are assumed to follow the exponential distribution. For this type of distribution, the probability of electric neutral strings is the highest, which corresponds to uncharged cosmic strings. The probability for cosmic strings with larger current becomes lower, which indicates that SCSs can hardly be formed at extremely high energy scales. Such an exponential distribution considered in the present work is not from a particular fundamental theory, but generally from an intuitive observation that, the lifetime of a string with the fixed tension should be shorter if it would carry a larger current due to the energy loss of both gravitational and electromagnetic radiations. We also update the constraints on the parameter space of SCSs using the latest published data.

The structure of this article is organized as follows. In Sec. \ref{S2} we depict the parameterized model of SCSs and the associated radiation mechanism. Then, in Sec. \ref{S3} we present the numerical estimation based on the newly proposed parametrization and report the updated constraint on the parameter space of these cosmic strings. Sec. \ref{S4} is devoted to the summary of our results accompanied by a discussion. In the theoretical derivations, we used the natural units with $\hbar = c = 1$. In the numerical calculations, we followed the setups of the Parkes multi-beam receiver and  ASKAP survey\footnote{We refer to \url{http://frbcat.org/} for details. Additionally, we have considered two data sets from radio experiments, which are the Parkes multi-beam system and ASKAP with redshift $z$ smaller than $2.1$ and $0.84$, respectively. All these events acquire extragalactic origin and the flux ranges within a few Jy.}.

\section{Characteristics of cosmic strings}
\label{S2}

Cosmic strings could form via phase transitions of the very early universe when the manifold of the vacuum background has nontrivial topology for symmetry breaking. These cosmic relics  reveal crucial information about fundamental physics at extremely high energy scales. In particular, these strings could carry a current, i.e. SCSs, which could be achieved by a charged scalar field whose flux is trapped in normal cosmic strings with the electromagnetic gauge invariance broken inside the strings, and hence would give rise to electromagnetic effects. Namely, as described in \cite{Ye:2017lqn}, radiations can be emitted from various structures of SCSs. In the following, we briefly review the radiation mechanism of SCSs and put forward a parametrized distribution function for SCSs.

\subsection{Parametrization of SCSs}

Recall that a cosmic string in general can emit both gravitational and electromagnetic radiation if there is a current. In the following context, we consider a string with the initial length $L_i$ at the initial time $t_i$. its length evolves as \cite{Vilenkin:2000jqa}
\begin{equation}
\label{L1}
 L(t) = L_i - \Gamma G{\mu}(t-t_i) ~,
\end{equation}
where $\Gamma$ is a scaling factor for the total radiation power $P$ from SCSs. Its form takes~\cite{Ye:2017lqn}
\begin{equation}
\label{Gamma}
 \Gamma = \frac{P}{G\mu^2} \simeq \frac{P_g+P^{c}_{\gamma}}{G\mu^2} ~,
\end{equation}
where $P_g$ and $P^{c}_{\gamma}$ denote the powers of gravitational and the cusp-sourced electromagnetic radiations, respectively.

As cosmic strings yield both gravitational and electromagnetic radiations, the lifetime of a typical string loop can be estimated by \cite{Vilenkin:2000jqa}
\begin{equation}
 \tau = \frac{L_i}{\Gamma G \mu} ~.
\end{equation}

Note that, in the literature on the electromagnetic effects of cosmic string, it was often assumed that all strings carry the same value of the current. However, this assumption could be relaxed by adopting a probability distribution for the currents inside various strings that may exist in the universe. Accordingly, the number density of SCSs can be expressed as
\begin{equation}
 dn_{SCS}=\Phi(I) ~ dn_{CS} ~ dI ~,
\end{equation}
where $dn_{SCS}$ is the number density of SCSs and $dn_{CS}$ is that of non-superconducting cosmic strings, while $\Phi(I)$ is the probability distribution for the current. In the following context, we discuss a toy model where $\Phi(I)$ is an exponential distribution:
\begin{equation}
\label{dis}
 \Phi(I)=\frac{1}{I_c}e^{-I/I_c} ~,
\end{equation}
with $I_c$ being a positively valued parameter. The choice of the exponential distribution is based on an intuitive picture that strings with smaller currents emit lower electromagnetic radiation power and hence will exist longer.

We comment that, as these topological defects have formed at different energy scales, the value of the corresponding current inside the strings varies along with the relevant energy scales for phase transitions. For infinite current, the probability approaches to zero, which is in agreement with the description of the exponential distribution function. Note that a truncated Gaussian distribution could be another attractive candidate for parametrization, in particular if one takes into account the possible correlation between the current and the string tension. In this case, there is automatically a theoretical bound on the maximal value of current for a fixed $G\mu$ when the current is uniformly distributed around a mean value for different strings. However, this case would be sensitive to the underlying theoretical models. As the scope of the present work focuses on the development of the analysis method on confronting the SCSs with FRB data, we wish to leave more detailed discussions on the parametrization of the distribution function in the future study.

\subsection{Radiations from string loops}

The gravitational radiation from cosmic strings can be expressed as \cite{Vilenkin:2000jqa}
\begin{equation}
 P_{g} = \Gamma_g G\mu^2~,
\end{equation}
where $\Gamma_g\sim100$. For electromagnetic radiation, the case becomes complicated. Note that, it is crucial to study the energy emitted from SCSs by taking into account the effect of the charge carriers upon the string motion itself, which has been exactly examined in the case of the chiral current in \cite{BlancoPillado:2000xy}. For a generic case, the electromagnetic backreaction becomes important when it is comparable with the string tension itself, and thus, this leads to a cutoff on the frequency of the emitted radiation as addressed in \cite{Cai:2012zd}, which is typically above the observational region for the FRB astronomy. The electromagnetic radiation power of SCS loops with length L mainly comes from cusps and kinks regions, which can be estimated as \cite{Vilenkin:1986zz, Cai:2012zd, Ye:2017lqn}
\begin{align}
\label{eqP}
 P_\gamma^c & \sim \kappa I \sqrt{\mu} ~, \\
 P_\gamma^k &\sim I^2 N \Psi \ln \biggl( \frac{\omega_{max}}{\omega_{min}} \biggr) ~,
\end{align}
where superscripts $c$ and $k$ stands for cusps and kinks, respectively. Here $\kappa\sim10$ is a numerical coefficient, $\omega_{max}$ and $\omega_{min}$ denote the highest and lowest frequency cutoffs for kinks, which are estimated as $\omega_{max} \sim \sqrt{\mu}$, and $\omega_{min} \sim (L/N)\sim 1$. We assume neither the numbers of kinks per loop N nor sharpness of discontinuity $\Psi$ would induce an order-of-magnitude change to the radiations from kinks, hence we take $N\Psi=1$ in this work.

\subsection{Event Rate of Burst}

%\textbf{We consider the signal from string network that reach scaling regime in the matter dominated epoch in which the loop distribution function has two components. The first component is established by signals from string loops that were born in the radiation dominated epoch but still outlasted in matter-dominated epoch and the second component is based on the loops that were born during the matter-dominated epoch. As a whole, the number density of loops produced in the matter-dominated era is the sum of these two components.} 

We consider the event rate of radio signals from string loops that reach scaling in the matter-dominated epoch.
Assuming the spatially flat Friedmann-Lemaitre-Robertson-Walker (FLRW) universe, the number density of these string loops can be expressed in terms of the redshift $z$ and the length $L$ as \cite{Cai:2012zd, Vilenkin:2000jqa}, 
\begin{equation}
 dn_{SCS}(z,L,I) \simeq \frac{C_L(z)(1+z)^6 \Phi(I)}{t_0^{2}[(1+z)^{3/2} L + \Gamma G\mu t_0 ]^2} dL dI ~,
\end{equation}
where the exponential distribution has already been considered and $t_0$ denotes the age of the Universe at present. Moreover, there is
\begin{equation}
 C_L(z)=1+{\frac{t_{eq}^{1/2}(1+z)^{3/4}}{\sqrt{(1+z)^{3/2}L+ \Gamma G\mu t_0}}} ~.
\end{equation}

The burst event rate in terms of string loop length $L$, redshift $z$, and kink sharpness $\Psi$, with the beam width $\Theta_w=(L \omega)^{-1/3}$ per unit volume is \cite{Cai:2012zd}:
\begin{equation}
\label{N1}
 d \dot {\mathcal{N}}(z,L,I) \simeq \frac{N^p \Theta_{\nu_0}^{-3m}}{L (1+z)} dn_{SCS}(z,L,I) dV(z) ~,
\end{equation}
where we have ($p = 0$, $m = -\frac{2}{3}$) for cusps and ($p = 1, m = -\frac{1}{3}$) for kinks. Additionally, there is $\Theta_{\nu_0} = [\nu_0 (1+z) L]^{-1/3}$ with $\nu_0$ being the observed frequency of the bursts. The physical volume element is given by,
\begin{equation}
 dV(z)= {4\pi}\frac{r_0(z)^2}{(1+z)^3}dz ~,
\end{equation}
where
\begin{equation}
 r_0(z)=3t_0\biggl(1-\frac{1}{\sqrt{1+z}}\biggr)~,
\end{equation}
is the comoving distance.

So far, we have demonstrated bursts event rate in terms of loop length and redshift. From an observational point of view, it is rather peculiar that we can relate the event bursts rate to measured variables, namely, the energy flux per frequency interval $S$ and the observed duration $\Delta$ of the burst. The burst event rate in terms of measured  parameters then takes the form \cite{Ye:2017lqn}:
\begin{equation}
\label{ENF}
 d \dot {\mathcal{N}}(S,z,I)=\tilde{A}\frac{[ {\nu}_0 L(S,z,I)]^m}{S}f_m(z,S,I) \Phi(I) dz dS dI ~,
\end{equation}
where,
\begin{align}
 & \tilde{A}=\frac{3 A N^p t_0}{2(1-3m)} ~, \\
 & L(z,S,I) = \biggl[ \frac{r_0(z)^2(1+z)^2 S \Delta}{I^2 \Psi^p} \biggr]^{\frac{3}{2(1-3m)}} \big[ \nu_0(1+z) \big]^{\frac{2+3m}{1-3m}} ~, \\
 & f_m(z,S,I)=\frac{C_L(z)(1+z)^{m-\frac{1}{2}}(\sqrt{1+z}-1)^2}{[(1+z)^{\frac{3}{2}}L + \Gamma G\mu t_0]^2} ~,
\end{align}
with $A\sim 50$. \cite{Martins:2005es,Vanchurin:2005pa,Ringeval:2005kr}

A radio signal, all the way from source to receiver, experiences scattering effect in various ways. For the observed width $\Delta$ of FRB signal, we considered both intrinsic and scattering duration caused by astrophysical medium given by \cite{Caleb:2015uuk}
\begin{equation}
 \Delta^2=\Delta t_{ISM}^2+\Delta t_{IGM}^2+\Delta t_{int}^2 ~,
\end{equation}
where $\Delta t_{ISM}$ and $\Delta t_{IGM}$ are scattering duration due to interstellar medium (ISM) and intergalactic medium (IGM) respectively. Note that, the ISM and IGM effects mainly affect the propagations of photons but not the sources when they are extragalactic. However, it is interesting to examine SCSs within IGM environment, such as involving certain intergalactic magnetic fields \cite{Gruzinov:2016hqs}. This may give rise to novel phenomena of experimental interest at different observational windows.

Due to ISM, the time broadening effects (scattering) for radio pulse, therefore yield \cite{Bhat:2004xt, Caleb:2015uuk}:
\begin{align}
 \log_{10}(\Delta t_{ISM}) = & -6.5+0.15\log_{10}(DM_{ISM}) \\
 & + 1.1\log_{10}(DM_{ISM})^2 - 3.9 \log_{10}(\nu_0) ~, \nonumber
\end{align}
where $DM_{ISM}$ is a constant equivalent to $95pc/cm^3$. Likewise, the rescaling of time broadening effect through IGM gives \cite{Caleb:2015uuk, Lorimer:2013roa}
\begin{align}
 \log_{10}(\Delta t_{IGM}) = & -9.5+0.15\log_{10}(DM_{IGM}) \\
 & + 1.1\log_{10}(DM_{IGM})^2 - 3.9 \log_{10}(\nu_0) ~. \nonumber
\end{align}
The observed FRBs are all from extra-galactic sources. Then, one finds that the dispersion measure of the IGM takes \cite{Deng:2013aga, Zhou:2014yta, Gao:2014iva}
\begin{equation}
 DM_{IGM}(z)=\frac{3cH_0\Omega_bf_{IGM}}{8\pi Gm_p}\int_{0}^{z}\frac{(1+z^\prime)dz^\prime}{E(z^\prime)} ~,
\end{equation}
with
\begin{equation}
 E(z)=\sqrt{\Omega_m(1+z)^3+\Omega_\Lambda} ~.
\end{equation}
where we have introduced the Hubble parameter $H_0$, the density parameter of matter {$\Omega_m$ }and that of dark energy {$\Omega_\Lambda$} of the present universe $\Omega_b$ is the baryon mass fraction of the universe, $f_{IGM}$ the fraction of baryon mass in the intergalactic medium, and $m_p$ is the mass of proton.

To deal with time dilation at observation point for cusps, one needs to consider the cosmological expansion factor.  Therefore, the intrinsic duration at point of observer is  \cite{Vilenkin:1986zz, Cai:2012zd, Ye:2017lqn}
\begin{equation}
 (\Delta_{int})_{cusp} \approx\frac{(1+z)L^{2/3}}{\nu_e^{1/3}} ~,
\end{equation}
where $\nu_e=\nu_0(1+z)$ is the emitted frequency of the radiation and ${\nu}_0$ is the observed one. The intrinsic duration of kinks is very short and is given by
\begin{equation}
 (\Delta_{int})_{kink}\approx\frac{1+z}{\nu_e} \sim 0 ~.
\end{equation}
Hence only the scattering effect will be considered.

\section{Numerical results}\label{S3}

At this stage, we have tackle the theoretical predictions of SCSs according to observational data of FRB. In this regard, there are five parameters in total that describe FRB data: $G\mu$, $I_c$, $\nu_0$, $\Delta$ and $S$. We fit the event burst rate to the normalized observed data, hence for each $(G\mu,I_c)$, we express the event burst rate in terms of redshift. As a result, we infer values for the pairs $(G\mu, I_c)$ according to FRB signals. For each $I_c$, the range of current $I$ is chosen to be $(0, 10^4]$GeV, based on the previous work \cite{Ye:2017lqn}, we need to integrate Eq. (\ref{ENF}) over respective bandwidth and energy flux for different receivers to get burst event rate for cusps and kinks. For Parkes and ASKAP receivers, using the radiometer equation, the threshold flux is \cite{Caleb:2015uuk, Bera:2016qfe},
\begin{align}
 S_*=\frac{SNR\times T_{sys}}{G_{sys}\sqrt{\Delta BN_{pol}}} ~,
\end{align} 
where $T_{sys}$ is the temperature of the system, $G_{sys}$ is the system gain, $B$ is bandwidth, $N_{pol}$ is the polarization number and $SNR$ is the signal-to-noise ratio. Following Table~\ref{table1}, one can get threshold flux for Parkes and ASKAP.

\begin{table}
\caption{Parameters of Parkes multibeam and ASKAP recivers. The values of system parameters including system gain $G_{sys}$, the bandwidth B, the polarization number $N_{pol}$, the system temperature $T_{sys}$ and the signal-to-noise ratio SNR are listed. }
\label{table1}
\begin{tabular}{||c| c c||}
\hline
& Parkes & ASKAP \\ [0.5ex]
\hline
$G_{sys}$(K/Jy) & 0.69& 0.029 \\
\hline
$B$(GHz)& 0.34 & 0.336 \\
\hline
$N_{pol}$ & 2 & 2 \\
\hline
$T_{sys}$(K) & 28 & 58 \\
\hline
SNR & 10 & 10 \\
\hline
\end{tabular}
\end{table}

\begin{table}
\caption{Observational parameters including redshift, energy flux and observed frequency for Parkes and ASKAP are listed.}
\label{table2}
\begin{tabular}{||c| c c||}
\hline
& Parkes & ASKAP \\ [0.5ex]
\hline
$z$ & [0, 2.1] & [0, 0.84] \\
\hline
$S$ (Jy)& $[10^{-1}, 10^{-2}]$ & $[10^{-1}, 10^{-2}]$ \\
\hline
$v_0$ (GHz) &  [1.182, 1.522] & [1.129, 1.465] \\
\hline
\end{tabular}
\end{table}

The observational parameters are displayed in Table~\ref{table2}. For the event rate of bursts, the contribution of flux has been suppressed outside the given range and is consistent with the detected events\cite{Ye:2017lqn}. We put few limitations on loop length for cusps. Given the tension of the string, from Eq. \eqref{eqP} we can get the upper bound on L, i.e. $L<\mu^{\frac{3}{2}}/(I^3\omega_{max})$. Also, for the given $G\mu$, one expects that the current through strings greater than a critical value, i.e. $I_*\simeq 10^{20}\times (G \mu)^{3/2}$ GeV yields a major contribution in the form of electromagnetic radiation \cite{Ye:2017lqn}. For statistical analysis, we divide each data set into 6 bins. Note that we normalize the data as:
\begin{equation}
 \sum y_{obs}\Delta z_{bin} = 1 ~,
\end{equation}
with $y_{obs}$ being the normalized event number per redshift. To examine the compatibility of theoretical data with the observed ones, we create different values of normalized event burst rate in terms of ($G\mu, I_c$) so that 
\begin{equation}
 \int y_i dz=1 ~.
\end{equation}
In order to quantify the ability of FRB observations to constrain the parameter space for SCSs, we perform the following $\chi^2$ fit:
\begin{equation}
 \chi^2=\sum_i^n \frac{(y_{obs} -y_i)^2}{e_{obs}^2} ~,
\end{equation}
where $n$ is the number of bins, $e_{obs}$ is the respective error bar of observational data for each redshift bin, and $y_i$ is the theoretical event rate at the center of redshift bin. We are interested in regime with bursts event rate per year between $10^2$ to $10^6$. 

\begin{table}
\caption{Best fits for Parkes and ASKAP with corresponding parameter values, $\dot{\mathcal{N}}$ per year and $\chi^{2}_{min}$ are provided.}
\label{table3}
\begin{tabular}{||c| c|c||}
\hline
& Parkes & ASKAP \\ [0.5ex]
\hline
$G\mu$ & $3.3 \times 10^{-14}$  & $1.24 \times 10^{-13}$ \\
\hline
$I_c$  (GeV)& $6.07$ & $2.41 \times 10^{-1}$ \\
\hline
$\chi^{2}_{min}$ & $0.77$ & $4.34$ \\
\hline
$\dot{\mathcal{N}}$ (yr$^{-1}$) & $2.2 \times 10^5$ & $1.94 \times 10^{3}$ \\
\hline
\end{tabular}
\end{table}
	
\begin{figure}
\includegraphics[width=0.45\textwidth]{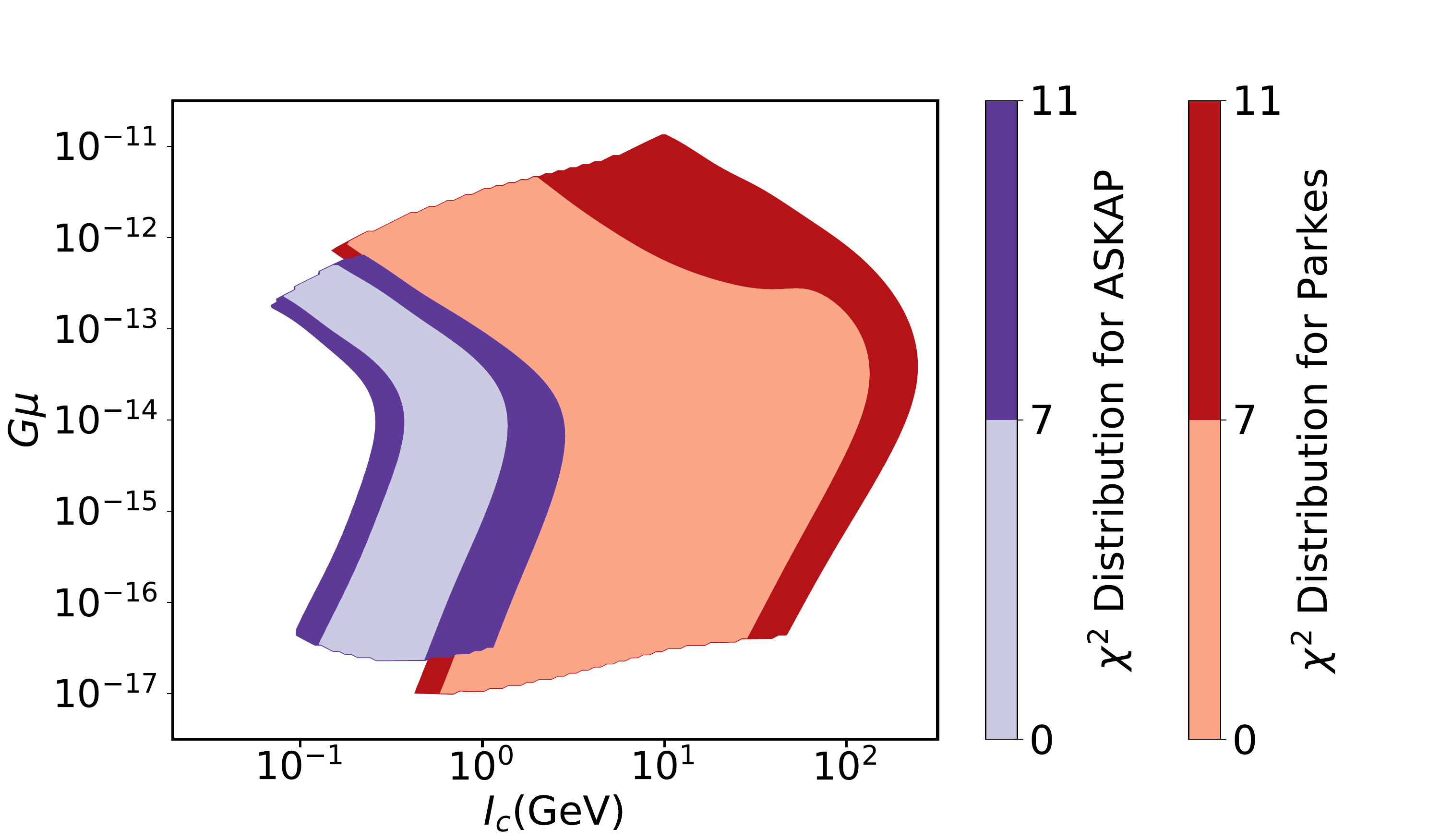}
\caption{The $\chi^2$ distribution with $2\sigma$ significance for Parkes (red shadow) and ASKAP (purple shadow), respectively. }
\label{fig:chi}
\end{figure}
	
\begin{figure}
\includegraphics[width=0.45\textwidth]{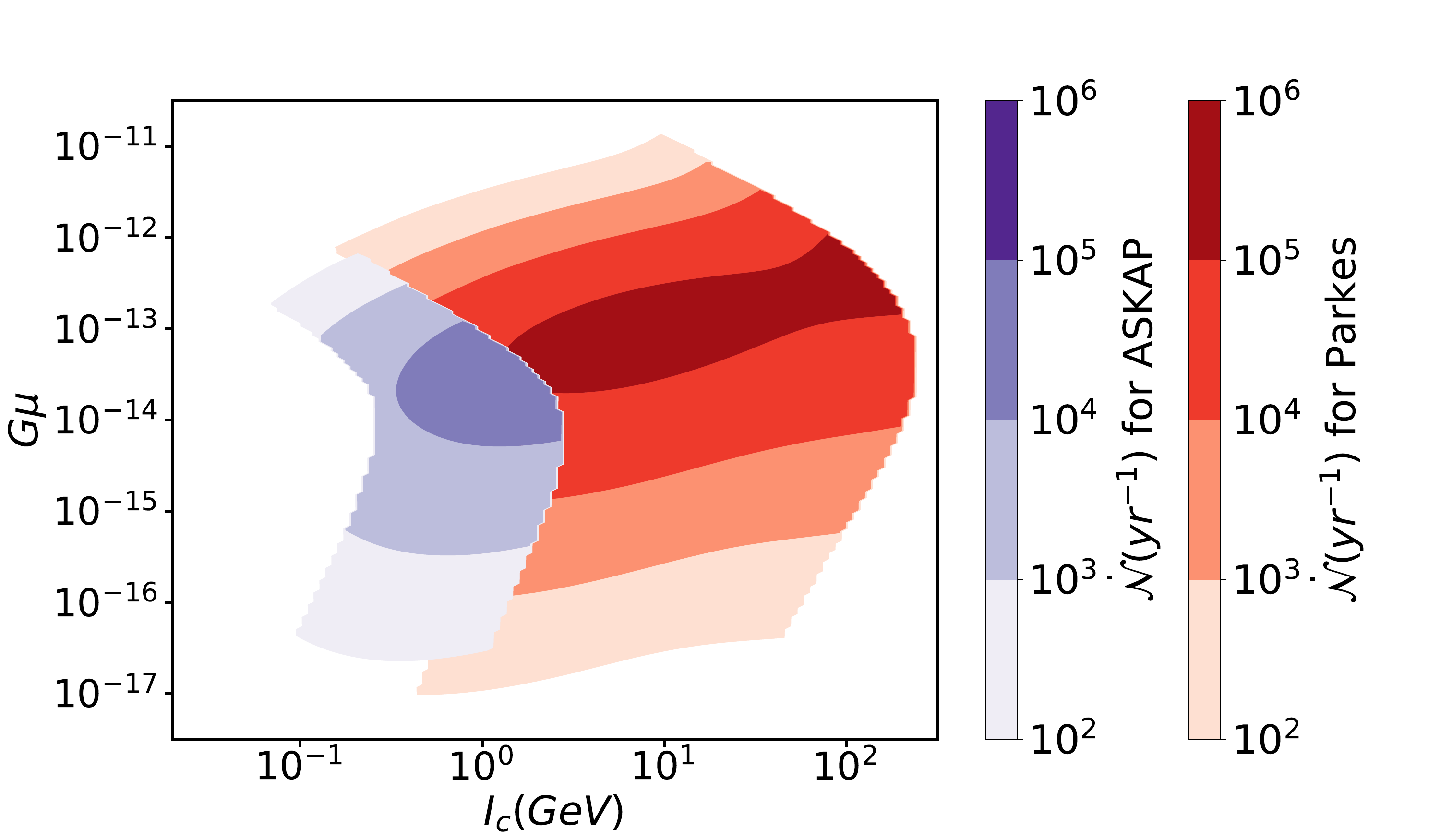}
\caption{The contour for the burst event rate distribution with $2\sigma$ significance for Parkes (red shadow) and ASKAP (purple shadow). The lighter to darker shadows of each colour correspond to the regimes with $\dot{\mathcal{N}}(yr^{-1})$ from $10^2$ to $10^6$, respectively. }
\label{fig:EN}
\end{figure}

\begin{figure}
\includegraphics[width=0.45\textwidth]{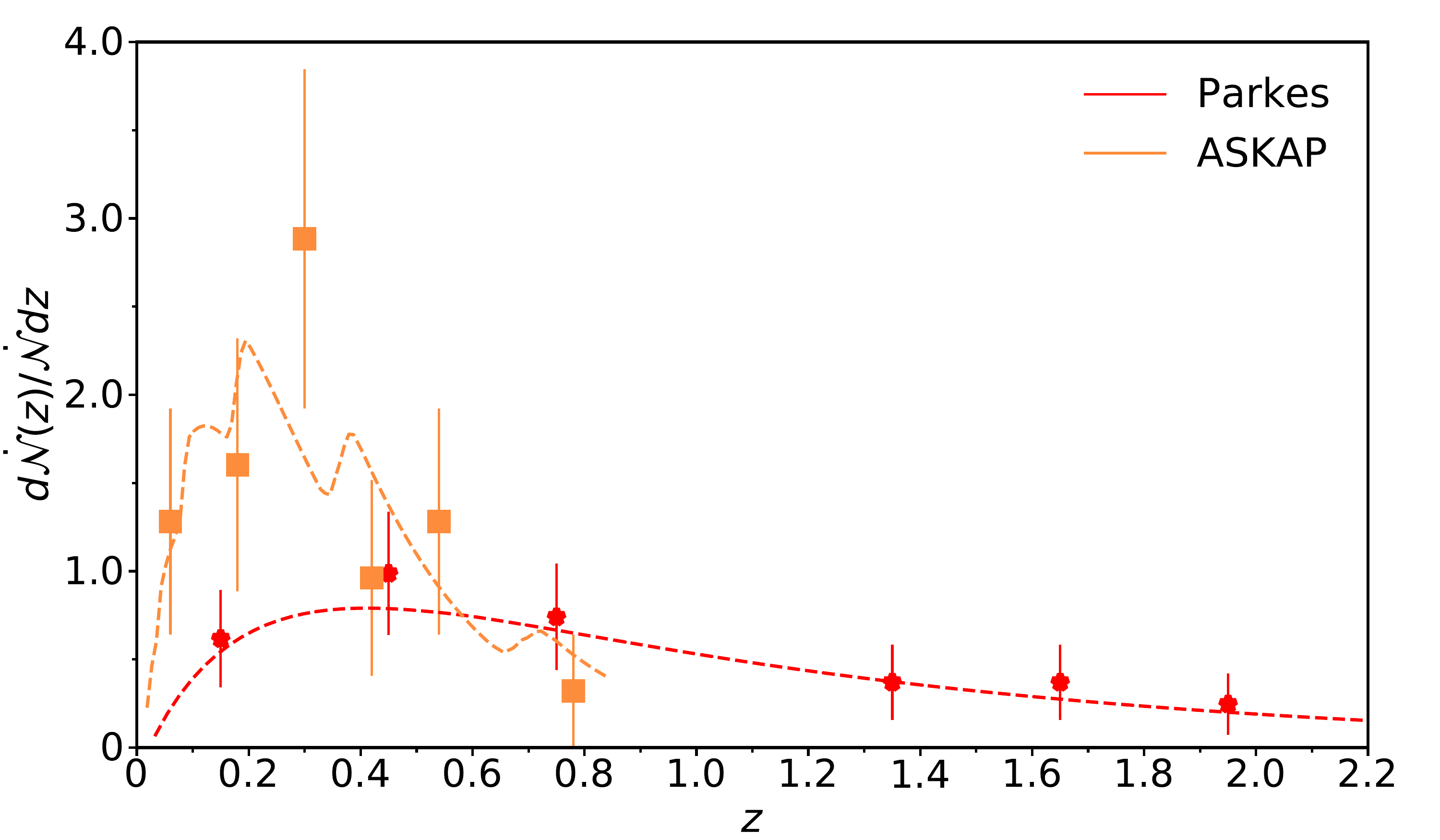}
\caption{The normalized observed event rate per redshift from Parkes (red) and ASKAP (orange), and theoretical fitting curve using SCSs model are shown with dashed lines. The observational parameters are listed in Table~\ref{table2}. The $\chi^2_{min}$ are $0.77$ for Parkes and $4.34$ for ASKAP. The corresponding parameter values with  event rates per year are listed in Table~\ref{table3}. }
\label{fig:fit}
\end{figure}

We investigate two observational data sets consisting of 25 signals from Parkes and 26 from ASKAP. In our analysis, we examine a $2\sigma$ significance level to scrutinize the parameter space for each experiment. For both instruments,  the significance of $2\sigma$ demonstrates the contour with $\chi^2<11$. This is shown in Fig.~\ref{fig:chi}. The red shaded region corresponds to Parkes and the purple for ASKAP. Both datasets allow us to constrain the parameter space to $G\mu\sim [10^{-17}, 10^{-12}]$ and $I_c \sim[10^{-1}, 10^2]$GeV. Further observations reveal that, for Parkes, the range for parameter space is slightly broadened out as calculated by \cite{Ye:2017lqn}, which could be due to the increase in both data points and redshift range. 

For the aforementioned estimated parameter space, we can infer the distribution of bursts event rate per year. Fig.~\ref{fig:EN} displays the findings of our analysis. We have calculated the event burst rate per year for each observational data set. The results of our analysis compatible with the ones of \cite{Rajwade:2016klq}. We calculate $\dot{\mathcal{N}}$ per year varies within the regime $10^2$ to $10^6$ and show that parameter space with string tension $G\mu\sim 10^{-14}$ gives the highest event rate per year. If SCSs radiate bursts with observable frequency $\nu_0$ and energy flux mentioned in Table~\ref{table3}, the burst production rate is the order of $10^3$-$10^5$ per year. Our results suggest that current and future radio experiments can probe SCS properties in this promising parameter space. For each  instrument, we notice that when $I_c>10^2$ GeV, the event number becomes stable for the corresponding string tension.

For the best fitting contours for each dataset in the estimated regime, we examine the best fits shown in Fig.~\ref{fig:fit}. In this figure, we compare the normalized observed event rate to our theoretical prediction. The model parameters corresponding to $\chi^{2}_{min}$ are given in Table~\ref{table3}. The best-fitting spot corresponding to $\dot{\mathcal{N}}$ per year for ASKAP is approximately consistent with the analysis done by \cite{Lu:2019pdn}. However, for Parkes, it might be quite lower than the estimated rate. In order to get an insight into further details, one may refer to \cite{Caleb:2015uuk}.

\section{Conclusion and discussion}
\label{S4}

To summarize, in the present article we have derived the updated constraints on the parameter space for SCS by confronting the theoretically predicted event rate with observational data of FRBs. We have analyzed the observed data from two telescopes, the Parkes and ASKAP.

Note that we have assumed all cosmic strings to be superconducting and the currents through the string loops follow the exponential distribution. We put forward a model in which the number density for SCSs can be defined in terms of number density for cosmic strings and the distribution of strings' currents. We notice the parameter space suggested by Parkes data for SCS suggested by \cite{Ye:2017lqn} is slightly broadened, demonstrated by Fig.~\ref{fig:chi}.

Another important finding is that the combination of Parkes and ASKAP constrained the parameter space well enough. Fig.~\ref{fig:chi} shows that the updated parameter space, allowed by FRB data from two radio experiments, can be constrained within: $G\mu \sim [10^{-17}, 10^{-12}]$ and $I_c \sim [10^{-1}, 10^2]$GeV. For each dataset, our detailed analysis declared a best-fitting contour estimated by $\chi^{2}_{min}$. The burst event rate per year for the aforementioned constrained parameter space ranges approximately between $10^2$ to $10^6$, which is consistent with the analysis conducted by \cite{Lu:2019pdn, Caleb:2015uuk}. 

The extant studies have been conducted under some simplifying assumptions that can be reconsidered in the future research. The string tension might rely on the current distribution, i.e. $\tilde{G\mu}=G\mu\times\Phi(I)$, while we have taken $G\mu$ to be the averaged tension along the string length in our work. Also, we have ignored the cosmological expansion for the number density defined in the dark energy era. As our universe has entered into a dark energy epoch, it is challenging to probe the radiation mechanism. The lack of convincing theoretical arguments to determine number density in the dark energy era has given rise to a discrepancy for this approach.
	
In addition, from all the detected FRBs, six repeated bursts from FRB180814.J0422+73 have been reported in \cite{Amiri:2019bjk}. We intend to look at two aspects for repeated bursts, one is observational and the other theoretical. For the observational aspect, we expect more data and on the theoretical line, we need more in-depth knowledge to explore these signals. In this regard, the precise measurement of energy released from our proposed parameterization could help to identify the theoretical origin for repeated FRBs.
	
Because of the precise constraints from FRBs, SCSs can be treated as the possible source for FRBs. In this perspective, the current in string loops could be in the range of GeV scale which makes SCS a useful tool to investigate high-energy paradigm along with collider experiments, for instance, the International Linear Collider, the Large Hadron Colloder and so forth.

\section*{Acknowledgement} 
The authors thank Zuyi Chen, Canmin Deng, Reinoud Slagter, Jiani Ye, Pierre Zhang for valuable communications. 
BI is supported by China Scholarship Council for the Ph.D. Program (No. 2016GXYN54).
YFC is supported in part by the NSFC (Nos. 11722327, 11653002, 11961131007, 11421303), by the CAST-YESS (2016QNRC001), by the National Youth  Talents Program of China, and by the Fundamental Research Funds for Central Universities.
All numerics are operated on the computer clusters {\it Linda \& Judy} in the particle cosmology group at USTC.


\begin{thebibliography}{99}

%\cite{Lorimer:2007qn}
\bibitem{Lorimer:2007qn} 
D.~R.~Lorimer, M.~Bailes, M.~A.~McLaughlin, D.~J.~Narkevic and F.~Crawford,
%``A bright millisecond radio burst of extragalactic origin,''
Science {\bf 318}, 777 (2007)
%doi:10.1126/science.1147532
[arXiv:0709.4301 [astro-ph]].
%%CITATION = doi:10.1126/science.1147532;%%

%\cite{Petroff:2019tty}
\bibitem{Petroff:2019tty}
E.~Petroff, J.~W.~T.~Hessels and D.~R.~Lorimer,
%``Fast Radio Bursts,''
Astron.\ Astrophys.\ Rev.\  {\bf 27} (2019) no.1,  4
%doi:10.1007/s00159-019-0116-6
[arXiv:1904.07947 [astro-ph.HE]].
%%CITATION = doi:10.1007/s00159-019-0116-6;%%

		%\cite{Cordes:2019cmq}
		\bibitem{Cordes:2019cmq} 
		J.~M.~Cordes and S.~Chatterjee,
		%``Fast Radio Bursts: An Extragalactic Enigma,''
		Ann.\ Rev.\ Astron.\ Astrophys.\  {\bf 57}, 417 (2019)
		%doi:10.1146/annurev-astro-091918-104501
		[arXiv:1906.05878 [astro-ph.HE]].
		%%CITATION = doi:10.1146/annurev-astro-091918-104501;%%
		
		%\cite{Keane:2011mj}
		\bibitem{Keane:2011mj} 
		E.~F.~Keane, M.~Kramer, A.~G.~Lyne, B.~W.~Stappers and M.~A.~McLaughlin,
		%``RRATs: New Discoveries, Timing Solutions \& Musings,''
		Mon.\ Not.\ Roy.\ Astron.\ Soc.\  {\bf 415}, 3065 (2011)
		%doi:10.1111/j.1365-2966.2011.18917.x
		[arXiv:1104.2727 [astro-ph.SR]].
		%%CITATION = doi:10.1111/j.1365-2966.2011.18917.x;%%
		
		%\cite{Thornton:2013iua}
		\bibitem{Thornton:2013iua} 
		D.~Thornton {\it et al.},
		%``A Population of Fast Radio Bursts at Cosmological Distances,''
		Science {\bf 341}, no. 6141, 53 (2013)
		%doi:10.1126/science.1236789
		[arXiv:1307.1628 [astro-ph.HE]].
		%%CITATION = doi:10.1126/science.1236789;%%
		
		%\cite{Champion:2015pmj}
		\bibitem{Champion:2015pmj} 
		D.~J.~Champion {\it et al.},
		%``Five new fast radio bursts from the HTRU high-latitude survey at Parkes: first evidence for two-component bursts,''
		Mon.\ Not.\ Roy.\ Astron.\ Soc.\  {\bf 460}, no. 1, L30 (2016)
		%doi:10.1093/mnrasl/slw069
		[arXiv:1511.07746 [astro-ph.HE]].
		%%CITATION = doi:10.1093/mnrasl/slw069;%%
		
		%\cite{Bhandari:2017qrj}
		\bibitem{Bhandari:2017qrj} 
		S.~Bhandari {\it et al.} [ANTARES Collaboration],
		%``The SUrvey for Pulsars and Extragalactic Radio Bursts – II. New FRB discoveries and their follow-up,''
		Mon.\ Not.\ Roy.\ Astron.\ Soc.\  {\bf 475}, no. 2, 1427 (2018)
		%doi:10.1093/mnras/stx3074
		[arXiv:1711.08110 [astro-ph.HE]].
		%%CITATION = doi:10.1093/mnras/stx3074;%%
		
		%\cite{Spitler:2014fla}
		\bibitem{Spitler:2014fla} 
		L.~G.~Spitler {\it et al.},
		%``Fast Radio Burst Discovered in the Arecibo Pulsar ALFA Survey,''
		Astrophys.\ J.\  {\bf 790}, no. 2, 101 (2014)
		%doi:10.1088/0004-637X/790/2/101
		[arXiv:1404.2934 [astro-ph.HE]].
		%%CITATION = doi:10.1088/0004-637X/790/2/101;%%
		
		%\cite{Patel:2018ubs}
		\bibitem{Patel:2018ubs} 
		C.~Patel {\it et al.},
		%``PALFA Single-Pulse Pipeline: New Pulsars, Rotating Radio Transients and a Candidate Fast Radio Burst,''
		Astrophys.\ J.\  {\bf 869}, no. 2, 181 (2018)
		%doi:10.3847/1538-4357/aaee65
		[arXiv:1808.03710 [astro-ph.HE]].
		%%CITATION = doi:10.3847/1538-4357/aaee65;%%
		
		%\cite{Masui:2015kmb}
		\bibitem{Masui:2015kmb} 
		K.~Masui {\it et al.},
		%``Dense magnetized plasma associated with a fast radio burst,''
		Nature {\bf 528}, 523 (2015)
		%doi:10.1038/nature15769
		[arXiv:1512.00529 [astro-ph.HE]].
		%%CITATION = doi:10.1038/nature15769;%%
		
		%\cite{Bannister:2017sie}
		\bibitem{Bannister:2017sie} 
		K.~Bannister {\it et al.},
		%``The detection of an extremely bright fast radio burst in a phased array feed survey,''
		Astrophys.\ J.\  {\bf 841}, L12 (2017)
		%doi:10.3847/2041-8213/aa71ff, 10.3847/2041-8213
		[arXiv:1705.07581 [astro-ph.HE]].
		%%CITATION = doi:10.3847/2041-8213/aa71ff, 10.3847/2041-8213;%%
		
		%\cite{Macquart:2018rsa}
		\bibitem{Macquart:2018rsa} 
		J.~P.~Macquart, R.~M.~Shannon, K.~W.~Bannister, C.~W.~James, R.~D.~Ekers and J.~D.~Bunton,
		%``The Spectral Properties of the Bright Fast Radio Burst Population,''
		Astrophys.\ J.\  {\bf 872}, no. 2, L19 (2019)
		%doi:10.3847/2041-8213/ab03d6
		[arXiv:1810.04353 [astro-ph.HE]].
		%%CITATION = doi:10.3847/2041-8213/ab03d6;%%
		
		%\cite{Caleb:2017vbk}
		\bibitem{Caleb:2017vbk} 
		M.~Caleb {\it et al.},
		%``The first interferometric detections of Fast Radio Bursts,''
		Mon.\ Not.\ Roy.\ Astron.\ Soc.\  {\bf 468}, no. 3, 3746 (2017)
		%doi:10.1093/mnras/stx638
		[arXiv:1703.10173 [astro-ph.HE]].
		%%CITATION = doi:10.1093/mnras/stx638;%%
		
		%\cite{Farah:2018buz}
		\bibitem{Farah:2018buz} 
		W.~Farah {\it et al.},
		%``FRB microstructure revealed by the real-time detection of FRB170827,''
		Mon.\ Not.\ Roy.\ Astron.\ Soc.\  {\bf 478}, no. 1, 1209 (2018)
		%doi:10.1093/mnras/sty1122
		[arXiv:1803.05697 [astro-ph.HE]].
		%%CITATION = doi:10.1093/mnras/sty1122;%%
		
		%\cite{Amiri:2019qbv}
		\bibitem{Amiri:2019qbv} 
		M.~Amiri {\it et al.} [CHIME/FRB Collaboration],
		%``Observations of fast radio bursts at frequencies down to 400 megahertz,''
		Nature {\bf 566}, no. 7743, 230 (2019)
		%doi:10.1038/s41586-018-0867-7
		[arXiv:1901.04524 [astro-ph.HE]].
		%%CITATION = doi:10.1038/s41586-018-0867-7;%%
		
		%\cite{Platts:2018hiy}
		\bibitem{Platts:2018hiy} 
		E.~Platts, A.~Weltman, A.~Walters, S.~P.~Tendulkar, J.~E.~B.~Gordin and S.~Kandhai,
		%``A Living Theory Catalogue for Fast Radio Bursts,''
		Phys.\ Rept.\  {\bf 821}, 1 (2019)
		%doi:10.1016/j.physrep.2019.06.003
		[arXiv:1810.05836 [astro-ph.HE]].
		%%CITATION = doi:10.1016/j.physrep.2019.06.003;%%
		
		%\cite{Metzger:2017wdz}
		\bibitem{Metzger:2017wdz}
		B.~D.~Metzger, E.~Berger and B.~Margalit,
		%``Millisecond Magnetar Birth Connects FRB 121102 to Superluminous Supernovae and Long Duration Gamma-ray Bursts,''
		Astrophys.\ J.\  {\bf 841} (2017) no.1,  14
		%doi:10.3847/1538-4357/aa633d
		[arXiv:1701.02370 [astro-ph.HE]].
		%%CITATION = doi:10.3847/1538-4357/aa633d;%%
		
		%\cite{Brandenberger:2017uwo}
		\bibitem{Brandenberger:2017uwo}
		R.~Brandenberger, B.~Cyr and A.~V.~Iyer,
		%``Fast Radio Bursts from the Decay of Cosmic String Cusps,''
		arXiv:1707.02397 [astro-ph.CO].
		%%CITATION = ARXIV:1707.02397;%%
		
		%\cite{Deng:2018wmy}
		\bibitem{Deng:2018wmy} 
		C.~M.~Deng, Y.~Cai, X.~F.~Wu and E.~W.~Liang,
		%``Fast Radio Bursts From Primordial Black Hole Binaries Coalescence,''
		Phys.\ Rev.\ D {\bf 98}, no. 12, 123016 (2018)
		%doi:10.1103/PhysRevD.98.123016
		[arXiv:1812.00113 [astro-ph.HE]].
		%%CITATION = doi:10.1103/PhysRevD.98.123016;%%
		
		%\cite{Wang:2016dgs}
		\bibitem{Wang:2016dgs} 
		J.~S.~Wang, Y.~P.~Yang, X.~F.~Wu, Z.~G.~Dai and F.~Y.~Wang,
		%``Fast Radio Bursts from the Inspiral of Double Neutron Stars,''
		Astrophys.\ J.\  {\bf 822}, no. 1, L7 (2016)
		%doi:10.3847/2041-8205/822/1/L7
		[arXiv:1603.02014 [astro-ph.HE]].
		%%CITATION = doi:10.3847/2041-8205/822/1/L7;%%
		
		%\cite{Dai:2016qem}
		\bibitem{Dai:2016qem} 
		Z.~G.~Dai, J.~S.~Wang, X.~F.~Wu and Y.~F.~Huang,
		%``Repeating Fast Radio Bursts from Highly Magnetized Pulsars Travelling through Asteroid Belts,''
		Astrophys.\ J.\  {\bf 829}, no. 1, 27 (2016)
		%doi:10.3847/0004-637X/829/1/27
		[arXiv:1603.08207 [astro-ph.HE]].
		%%CITATION = doi:10.3847/0004-637X/829/1/27;%%
		
		%\cite{Zhang:2017zse}
		\bibitem{Zhang:2017zse} 
		B.~Zhang,
		%``A Cosmic Comb Model of Fast Radio Bursts,''
		Astrophys.\ J.\  {\bf 836}, no. 2, L32 (2017)
		%doi:10.3847/2041-8213/aa5ded
		[arXiv:1701.04094 [astro-ph.HE]].
		%%CITATION = doi:10.3847/2041-8213/aa5ded;%%
		
		%\cite{Palaniswamy:2017aze}
		\bibitem{Palaniswamy:2017aze} 
		D.~Palaniswamy, Y.~Li and B.~Zhang,
		%``Are there multiple populations of Fast Radio Bursts?,''
		Astrophys.\ J.\  {\bf 854}, no. 1, L12 (2018)
		%doi:10.3847/2041-8213/aaaa63
		[arXiv:1703.09232 [astro-ph.HE]].
		%%CITATION = doi:10.3847/2041-8213/aaaa63;%%
		
		%\cite{Vilenkin:2000jqa}
		\bibitem{Vilenkin:2000jqa} 
		A.~Vilenkin and E.~P.~S.~Shellard,
		%``Cosmic Strings and Other Topological Defects,''
		(Cambridge University Press, Cambridge, 1994).
		%%CITATION = INSPIRE-1384873;%%
		
		%\cite{Kibble:1976sj}
		\bibitem{Kibble:1976sj}
		T.~W.~B.~Kibble,
		%``Topology of Cosmic Domains and Strings,''
		J.\ Phys.\ A {\bf 9}, 1387 (1976).
		%doi:10.1088/0305-4470/9/8/029
		%%CITATION = doi:10.1088/0305-4470/9/8/029;%%
		
		%\cite{Witten:1984eb}
		\bibitem{Witten:1984eb} 
		E.~Witten,
		%``Superconducting Strings,''
		Nucl.\ Phys.\ B {\bf 249}, 557 (1985).
		%doi:10.1016/0550-3213(85)90022-7
		%%CITATION = doi:10.1016/0550-3213(85)90022-7;%%
		
		%\cite{Vilenkin:1986zz}
		\bibitem{Vilenkin:1986zz} 
		A.~Vilenkin and T.~Vachaspati,
		%``Electromagnetic Radiation from Superconducting Cosmic Strings,''
		Phys.\ Rev.\ Lett.\  {\bf 58}, 1041 (1987).
		%doi:10.1103/PhysRevLett.58.1041
		%%CITATION = doi:10.1103/PhysRevLett.58.1041;%%
		
		%\cite{Garfinkle:1987yw}
		\bibitem{Garfinkle:1987yw} 
		D.~Garfinkle and T.~Vachaspati,
		%``Radiation From Kinky, Cuspless Cosmic Loops,''
		Phys.\ Rev.\ D {\bf 36}, 2229 (1987).
		%doi:10.1103/PhysRevD.36.2229
		%%CITATION = doi:10.1103/PhysRevD.36.2229;%%
		
		%\cite{Vachaspati:2008su}
		\bibitem{Vachaspati:2008su} 
		T.~Vachaspati,
		%``Cosmic Sparks from Superconducting Strings,''
		Phys.\ Rev.\ Lett.\  {\bf 101}, 141301 (2008)
		%doi:10.1103/PhysRevLett.101.141301
		[arXiv:0802.0711 [astro-ph]].
		%%CITATION = doi:10.1103/PhysRevLett.101.141301;%%
		
		%\cite{Cai:2011bi}
		\bibitem{Cai:2011bi} 
		Y.~F.~Cai, E.~Sabancilar and T.~Vachaspati,
		%``Radio bursts from superconducting strings,''
		Phys.\ Rev.\ D {\bf 85}, 023530 (2012)
		%doi:10.1103/PhysRevD.85.023530
		[arXiv:1110.1631 [astro-ph.CO]].
		%%CITATION = doi:10.1103/PhysRevD.85.023530;%%
		
		%\cite{Cai:2012zd}
		\bibitem{Cai:2012zd} 
		Y.~F.~Cai, E.~Sabancilar, D.~A.~Steer and T.~Vachaspati,
		%``Radio Broadcasts from Superconducting Strings,''
		Phys.\ Rev.\ D {\bf 86}, 043521 (2012)
		%doi:10.1103/PhysRevD.86.043521
		[arXiv:1205.3170 [astro-ph.CO]].
		%%CITATION = doi:10.1103/PhysRevD.86.043521;%%
		
		%\cite{Ye:2017lqn}
		\bibitem{Ye:2017lqn} 
		J.~Ye, K.~Wang and Y.~F.~Cai,
		%``Superconducting cosmic strings as sources of cosmological fast radio bursts,''
		Eur.\ Phys.\ J.\ C {\bf 77}, no. 11, 720 (2017)
		%doi:10.1140/epjc/s10052-017-5319-2
		[arXiv:1705.10956 [astro-ph.HE]].
		%%CITATION = doi:10.1140/epjc/s10052-017-5319-2;%%
		
		
		
		%\cite{Yu:2014gea}
		\bibitem{Yu:2014gea}
		Y.~W.~Yu, K.~S.~Cheng, G.~Shiu and H.~Tye,
		%``Implications of fast radio bursts for superconducting cosmic strings,''
		JCAP {\bf 1411} (2014) no.11,  040
		%doi:10.1088/1475-7516/2014/11/040
		[arXiv:1409.5516 [astro-ph.HE]].
		%%CITATION = doi:10.1088/1475-7516/2014/11/040;%%
		
		%\cite{Dvorkin:2011aj}
		\bibitem{Dvorkin:2011aj} 
		C.~Dvorkin, M.~Wyman and W.~Hu,
		%``Cosmic String constraints from WMAP and the South Pole Telescope,''
		Phys.\ Rev.\ D {\bf 84}, 123519 (2011)
		%doi:10.1103/PhysRevD.84.123519
		[arXiv:1109.4947 [astro-ph.CO]].
		%%CITATION = doi:10.1103/PhysRevD.84.123519;%%
		
		%\cite{Ade:2013xla}
		\bibitem{Ade:2013xla} 
		P.~A.~R.~Ade {\it et al.} [Planck Collaboration],
		%``Planck 2013 results. XXV. Searches for cosmic strings and other topological defects,''
		Astron.\ Astrophys.\  {\bf 571}, A25 (2014)
		%doi:10.1051/0004-6361/201321621
		[arXiv:1303.5085 [astro-ph.CO]].
		%%CITATION = doi:10.1051/0004-6361/201321621;%%
		
		%\cite{vanHaasteren:2011ni}
		\bibitem{vanHaasteren:2011ni} 
		R.~van Haasteren {\it et al.},
		%``Placing limits on the stochastic gravitational-wave background using European Pulsar Timing Array data,''
		Mon.\ Not.\ Roy.\ Astron.\ Soc.\  {\bf 414}, no. 4, 3117 (2011)
		%Erratum: [Mon.\ Not.\ Roy.\ Astron.\ Soc.\  {\bf 425}, no. 2, 1597 (2012)]
		%doi:10.1111/j.1365-2966.2011.18613.x, 10.1111/j.1365-2966.2012.20916.x
		[arXiv:1103.0576 [astro-ph.CO]].
		%%CITATION = doi:10.1111/j.1365-2966.2011.18613.x, 10.1111/j.1365-2966.2012.20916.x;%%
		
		%\cite{Pshirkov:2009vb}
		\bibitem{Pshirkov:2009vb}
		M.~S.~Pshirkov and A.~V.~Tuntsov,
		%``Local constraints on cosmic string loops from photometry and pulsar timing,''
		Phys.\ Rev.\ D {\bf 81}, 083519 (2010)
		%doi:10.1103/PhysRevD.81.083519
		[arXiv:0911.4955 [astro-ph.CO]].
		%%CITATION = doi:10.1103/PhysRevD.81.083519;%%
		
		%\cite{Tuntsov:2010fu}
		\bibitem{Tuntsov:2010fu}
		A.~V.~Tuntsov and M.~S.~Pshirkov,
		%``Quasar variability limits on cosmological density of cosmic strings,''
		Phys.\ Rev.\ D {\bf 81}, 063523 (2010)
		%doi:10.1103/PhysRevD.81.063523
		[arXiv:1001.4580 [astro-ph.CO]].
		%%CITATION = doi:10.1103/PhysRevD.81.063523;%%
		
		%\cite{Olmez:2010bi}
		\bibitem{Olmez:2010bi}
		S.~Olmez, V.~Mandic and X.~Siemens,
		%``Gravitational-Wave Stochastic Background from Kinks and Cusps on Cosmic Strings,''
		Phys.\ Rev.\ D {\bf 81}, 104028 (2010)
		%doi:10.1103/PhysRevD.81.104028
		[arXiv:1004.0890 [astro-ph.CO]].
		%%CITATION = doi:10.1103/PhysRevD.81.104028;%%
		
		%\cite{Sanidas:2012ee}
		\bibitem{Sanidas:2012ee}
		S.~A.~Sanidas, R.~A.~Battye and B.~W.~Stappers,
		%``Constraints on cosmic string tension imposed by the limit on the stochastic gravitational wave background from the European Pulsar Timing Array,''
		Phys.\ Rev.\ D {\bf 85}, 122003 (2012)
		%doi:10.1103/PhysRevD.85.122003
		[arXiv:1201.2419 [astro-ph.CO]].
		%%CITATION = doi:10.1103/PhysRevD.85.122003;%%
		
		%\cite{Binetruy:2012ze}
		\bibitem{Binetruy:2012ze}
		P.~Binetruy, A.~Bohe, C.~Caprini and J.~F.~Dufaux,
		%``Cosmological Backgrounds of Gravitational Waves and eLISA/NGO: Phase Transitions, Cosmic Strings and Other Sources,''
		JCAP {\bf 1206}, 027 (2012)
		%doi:10.1088/1475-7516/2012/06/027
		[arXiv:1201.0983 [gr-qc]].
		%%CITATION = doi:10.1088/1475-7516/2012/06/027;%%
		
		%\cite{Sanchez:1988ek}
		\bibitem{Sanchez:1988ek}
		N.~G.~Sanchez and M.~Signore,
		%``The Cosmological Microwave Background Radiation, Cosmic and Superconducting Strings,''
		Phys.\ Lett.\ B {\bf 219}, 413 (1989).
		%doi:10.1016/0370-2693(89)91087-3
		%%CITATION = doi:10.1016/0370-2693(89)91087-3;%%
		
		%\cite{Tashiro:2012nb}
		\bibitem{Tashiro:2012nb}
		H.~Tashiro, E.~Sabancilar and T.~Vachaspati,
		%``CMB Distortions from Superconducting Cosmic Strings,''
		Phys.\ Rev.\ D {\bf 85}, 103522 (2012)
		%doi:10.1103/PhysRevD.85.103522
		[arXiv:1202.2474 [astro-ph.CO]].
		%%CITATION = doi:10.1103/PhysRevD.85.103522;%%
		
		%\cite{Acharya:2019xla}
		\bibitem{Acharya:2019xla} 
		S.~K.~Acharya and R.~Khatri,
		%``CMB spectral distortions constraints on primordial black holes, cosmic strings and long lived unstable particles revisited,''
		arXiv:1912.10995 [astro-ph.CO].
		%%CITATION = ARXIV:1912.10995;%%
		
		%\cite{Kogut:2011xw}
		\bibitem{Kogut:2011xw}
		A.~Kogut {\it et al.},
		%``The Primordial Inflation Explorer (PIXIE): A Nulling Polarimeter for Cosmic Microwave Background Observations,''
		JCAP {\bf 1107}, 025 (2011)
		%doi:10.1088/1475-7516/2011/07/025
		[arXiv:1105.2044 [astro-ph.CO]].
		%%CITATION = doi:10.1088/1475-7516/2011/07/025;%%
		
		%\cite{BlancoPillado:2000xy}
		\bibitem{BlancoPillado:2000xy} 
		J.~J.~Blanco-Pillado and K.~D.~Olum,
		%``Electromagnetic radiation from superconducting string cusps,''
		Nucl.\ Phys.\ B {\bf 599}, 435 (2001)
		%doi:10.1016/S0550-3213(00)00771-9
		[astro-ph/0008297].
		%%CITATION = doi:10.1016/S0550-3213(00)00771-9;%%
		
		%\cite{Martins:2005es}
		\bibitem{Martins:2005es}
		C.~J.~A.~Martins and E.~P.~S.~Shellard,
		%``Fractal properties and small-scale structure of cosmic string networks,''
		Phys.\ Rev.\  D {\bf 73}, 043515 (2006)
		%[arXiv:astro-ph/0511792].
		%%CITATION = PHRVA,D73,043515;%%
		
		%\cite{Vanchurin:2005pa}
		\bibitem{Vanchurin:2005pa}
		V.~Vanchurin, K.~D.~Olum and A.~Vilenkin,
		%``Scaling of cosmic string loops,''
		Phys.\ Rev.\  D {\bf 74}, 063527 (2006)
		%[arXiv:gr-qc/0511159];
		%%CITATION = PHRVA,D74,063527;%%
		
		%\cite{Ringeval:2005kr}
		\bibitem{Ringeval:2005kr}
		C.~Ringeval, M.~Sakellariadou and F.~Bouchet,
		%``Cosmological evolution of cosmic string loops,''
		JCAP {\bf 0702}, 023 (2007)
		%[arXiv:astro-ph/0511646];
		%%CITATION = JCAPA,0702,023;%%
		
		%\cite{Caleb:2015uuk}
		\bibitem{Caleb:2015uuk} 
		M.~Caleb, C.~Flynn, M.~Bailes, E.~D.~Barr, R.~W.~Hunstead, E.~F.~Keane, V.~Ravi and W.~van Straten,
		%``Are the distributions of Fast Radio Burst properties consistent with a cosmological population?,''
		Mon.\ Not.\ Roy.\ Astron.\ Soc.\  {\bf 458}, no. 1, 708 (2016)
		%doi:10.1093/mnras/stw175
		[arXiv:1512.02738 [astro-ph.HE]].
		%%CITATION = doi:10.1093/mnras/stw175;%%
		
		%\cite{Gruzinov:2016hqs}
		\bibitem{Gruzinov:2016hqs}
		A.~Gruzinov and A.~Vilenkin,
		%``Fireballs from Superconducting Cosmic Strings,''
		JCAP {\bf 1701} (2017) 029
		%doi:10.1088/1475-7516/2017/01/029
		[arXiv:1608.05396 [astro-ph.HE]].
		%%CITATION = doi:10.1088/1475-7516/2017/01/029;%%
		
		%\cite{Bhat:2004xt}
		\bibitem{Bhat:2004xt} 
		N.~D.~R.~Bhat, J.~M.~Cordes, F.~Camilo, D.~J.~Nice and D.~R.~Lorimer,
		%``Multifrequency observations of radio pulse broadening and constraints on interstellar electron density microstructure,''
		Astrophys.\ J.\  {\bf 605}, 759 (2004)
		%doi:10.1086/382680
		[astro-ph/0401067].
		%%CITATION = doi:10.1086/382680;%%
		
		%\cite{Lorimer:2013roa}
		\bibitem{Lorimer:2013roa} 
		D.~R.~Lorimer, A.~Karastergiou, M.~A.~McLaughlin and S.~Johnston,
		%``On the detectability of extragalactic fast radio transients,''
		Mon.\ Not.\ Roy.\ Astron.\ Soc.\  {\bf 436}, 5 (2013)
		%doi:10.1093/mnrasl/slt098
		[arXiv:1307.1200 [astro-ph.HE]].
		%%CITATION = doi:10.1093/mnrasl/slt098;%%
		
		
		%\cite{Deng:2013aga}
		\bibitem{Deng:2013aga} 
		W.~Deng and B.~Zhang,
		%``Cosmological Implications of Fast Radio Burst/Gamma-Ray Burst Associations,''
		Astrophys.\ J.\  {\bf 783}, L35 (2014)
		%doi:10.1088/2041-8205/783/2/L35
		[arXiv:1401.0059 [astro-ph.HE]].
		%%CITATION = doi:10.1088/2041-8205/783/2/L35;%%
		
		%\cite{Zhou:2014yta}
		\bibitem{Zhou:2014yta} 
		B.~Zhou, X.~Li, T.~Wang, Y.~Z.~Fan and D.~M.~Wei,
		%``Fast radio bursts as a cosmic probe?,''
		Phys.\ Rev.\ D {\bf 89}, no. 10, 107303 (2014)
		%doi:10.1103/PhysRevD.89.107303
		[arXiv:1401.2927 [astro-ph.CO]].
		%%CITATION = doi:10.1103/PhysRevD.89.107303;%%
		
		%\cite{Gao:2014iva}
		\bibitem{Gao:2014iva} 
		H.~Gao, Z.~Li and B.~Zhang,
		%``Fast Radio Burst/Gamma-Ray Burst Cosmography,''
		Astrophys.\ J.\  {\bf 788}, 189 (2014)
		%doi:10.1088/0004-637X/788/2/189
		[arXiv:1402.2498 [astro-ph.CO]].
		%%CITATION = doi:10.1088/0004-637X/788/2/189;%%
		
		%\cite{Bera:2016qfe}
		\bibitem{Bera:2016qfe} 
		A.~Bera, S.~Bhattacharyya, S.~Bharadwaj, N.~D.~R.~Bhat and J.~N.~Chengalur,
		%``On modelling the Fast Radio Burst population and event rate predictions,''
		Mon.\ Not.\ Roy.\ Astron.\ Soc.\  {\bf 457}, no. 3, 2530 (2016)
		%doi:10.1093/mnras/stw177
		[arXiv:1601.05410 [astro-ph.HE]].
		%%CITATION = doi:10.1093/mnras/stw177;%%
		
		%\cite{Rajwade:2016klq}
		\bibitem{Rajwade:2016klq}
		K.~Rajwade and D.~Lorimer,
		%``Detecting fast radio bursts at decametric wavelengths,''
		Mon.\ Not.\ Roy.\ Astron.\ Soc.\  {\bf 465} (2017) no.2,  2286
		doi:10.1093/mnras/stw2914
		[arXiv:1609.00929 [astro-ph.HE]].
		%%CITATION = doi:10.1093/mnras/stw2914;%%
		%18 citations counted in INSPIRE as of 14 Apr 2020
		
		%\cite{Lu:2019pdn}
		\bibitem{Lu:2019pdn}
		W.~Lu and A.~L.~Piro,
		%``Implications from ASKAP Fast Radio Burst Statistics,''
		doi:10.3847/1538-4357/ab3796
		arXiv:1903.00014 [astro-ph.HE].
		%%CITATION = doi:10.3847/1538-4357/ab3796;%%
		%9 citations counted in INSPIRE as of 15 Apr 2020
		
		%\cite{Amiri:2019bjk}
		\bibitem{Amiri:2019bjk} 
		M.~Amiri {\it et al.} [CHIME/FRB Collaboration],
		%``A Second Source of Repeating Fast Radio Bursts,''
		Nature {\bf 566}, no. 7743, 235 (2019)
		%doi:10.1038/s41586-018-0864-x
		[arXiv:1901.04525 [astro-ph.HE]].
		%%CITATION = doi:10.1038/s41586-018-0864-x;%%
		
		
	\end{thebibliography}
\end{document}